\documentclass[twocolumn, prl, superscriptaddress]{revtex4-1}
\usepackage{amssymb}
\usepackage{amsmath}
\usepackage{epsfig}
\usepackage{color}
\usepackage{graphics, graphicx}
\usepackage{bbold}
\usepackage{psfrag}
\usepackage{mathcomp}
\usepackage{subfigure}
\usepackage{verbatim}
\usepackage{color}
\usepackage{ulem}
\usepackage{upgreek}
\usepackage[colorlinks,citecolor=blue]{hyperref}

\begin{document}
\title{Isotope Separation of Potassium with a magneto-optical combined method}
\author{Zixuan Zeng}
\author{Shangjin Li}
\affiliation{%
Interdisciplinary Center of Quantum Information, State Key Laboratory of Modern Optical Instrumentation, and Zhejiang Province Key Laboratory of Quantum Technology and Device of Physics Department, Zhejiang University, Hangzhou 310027, China
}%
\author{Bo Yan}
\email{yanbohang@zju.edu.cn}
\affiliation{%
Interdisciplinary Center of Quantum Information, State Key Laboratory of Modern Optical Instrumentation, and Zhejiang Province Key Laboratory of Quantum Technology and Device of Physics Department, Zhejiang University, Hangzhou 310027, China
}%

\date{\today}

\begin{abstract}
Due to the similar physical and chemical properties, isotopes are usually hard to separate. On the other hand, the isotope shifts are very well separated in a high-resolution spectrum, making them possible to be addressed individually by lasers, thus separated. Here we report such an isotope separation experiment with Potassium atoms. The isotopes are independently optical pumped to the desired spin states, and then separated with a Stern-Gerlach scheme. A micro-capillary oven is used to collimate the atomic beam, and a Halbach-type magnet array is used to deflect the desired atoms. Finally, the $^{40}$K is enriched by two orders of magnitude. This magneto-optical combined method provides an effective way to separate isotopes and can be extended to other elements if the relevant optical pumping scheme is feasible.
\end{abstract}
\maketitle

Isotope separation has important applications in modern society. The most famous example is the Uranium isotope separation. Nowadays, nuclear power industry still has heavy requirement of isotope separation, except Uranium, high purity $^{10}$B \cite{Lyakhov2017}, $^{7}$Li \cite{Mazur2014} are indispensable parts in the nuclear power industry. Isotopes can also be used to trace important processes in environment and earth science \cite{HongMing2012,Bartelink2019}. Some isotopes, such as $^{13}$C, $^{15}$N, can be used to tracing the complex biological processes \cite{Wilkinson2018}. Besides, the isotope ratio measurement is proved an excellent way to perform dating analysis \cite{Muller1977, Chen1999}. In medical , agriculture,  manufacturing industry et al., important applications of isotope can be seen everywhere. Isotopes like a treasure trove, waiting us to unveil the nature mystery. But one thing hinders more extensive exploration of the applications, the high price of isotope separation. A convenient way to effectively separate isotopes is always warmly welcome.

Due to their similar physical and chemical properties, isotope separation is usually not easy. There are few routes to achieve this goal. One of the general methods is cyclotron which ionize atoms and separate them by different charge to mass ratios \cite{Yergey1997}. The cyclotron can provide a high enrich factor, and works for most isotopes \cite{Love1973}. However, such machine requires very large energy input. Other important methods for massive isotope production include the centrifuge separation and chemical separation, but they usually have very small enrich factors and multi-stage separations are needed. 
\begin{figure}[tbp]
\includegraphics[width=0.4\textwidth]{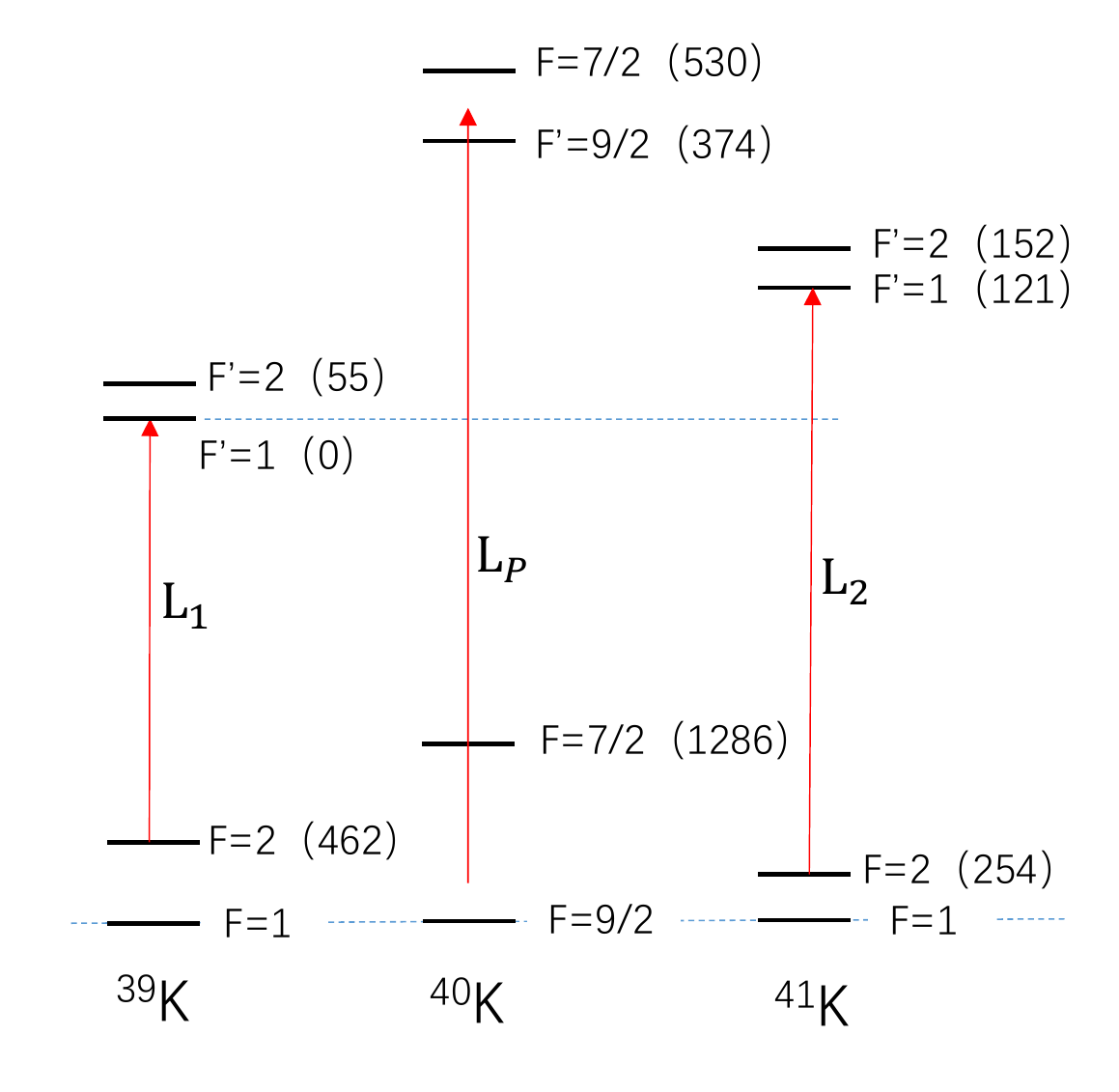}
\caption{\label{fig1}The energy levels of Potassium for different isotopes. The dash lines indicate the reference energy levels for the ground and excited states (D1 line). The number in the brackets shows the frequency offset (unit of MHz)  compared with the reference levels. $L_1$ and $L_2$ are the pumping lasers for $^{39}$K and $^{41}$K, and both are on resonant with $F=2$ to$F'=1$ transition. $L_P$ is the probe laser.}
\end{figure}

On the other hand, the high resolution spectrum of isotopes shows quite different features. The isotope shifts are on the same order of hyperfine splitting ( MHz to GHz regime). This value is small compared with the typical Doppler broadening, but much higher than the natural linewidth. With a sub-Doppler spectrum technique, different isotopes can be easily identified in the spectrum.
Thanks to the well-developed laser technique, we can easily get a laser with linewidth less than 1 MHz nowadays, which is much less than the isotope shifts, thus make it possible to address different isotopes individually.

\begin{figure*}[tbp]
\includegraphics[width=0.92\textwidth]{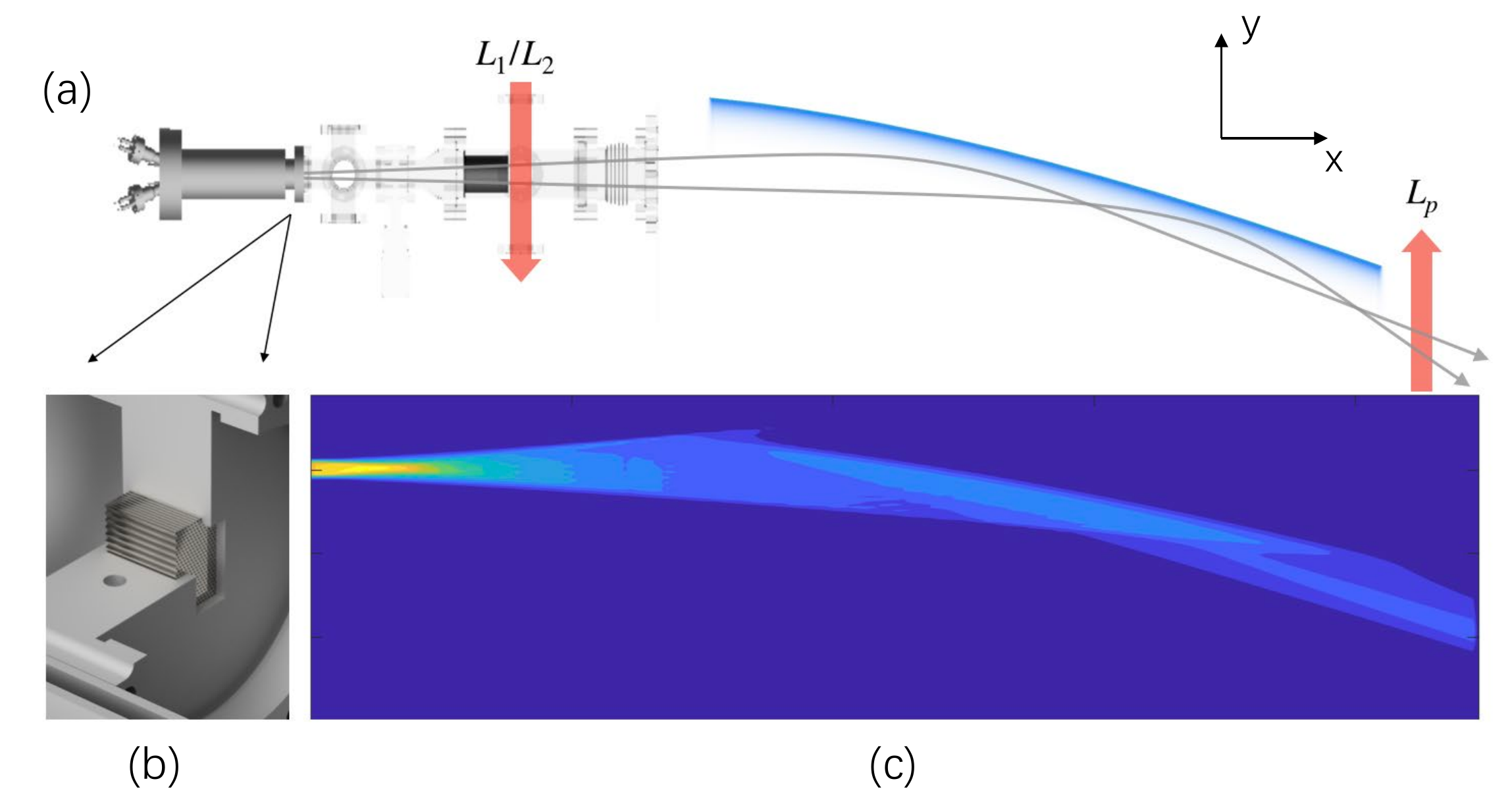}
\caption{\label{fig2}(a) The diagrammatic sketch for the Potassium isotope separation machine. Atoms are emitted from a heated oven and pumped by $L_1$ and $L_2$. A curved magnet array deflects atoms with the right magnetic moment. And finally atoms are imaged by the probe laser $L_p$.  (b)   In order to get a high flow rate, a micro-capillary array is used to collimate the atomic beam. (c) The typical trajectory of atoms in the magnetic field. Atoms with different incidence angles are deflected at different positions of the magnets.}
\end{figure*}

This leads to the idea of laser isotope separation, which has been extensively explored during 1980s to 90s \cite{Letokhov1979,ARISAWA1982, Paisner1988,GREENLAND1990}. At that time, the main strategy is to ionize atoms with laser. Due to the isotope shifts, the ionization efficiencies will be different for isotopes \cite{Kieck2019}, thus lead to separation. But with Doppler broadening and limited ionization efficiency, the total separation efficiency is not very high. Another idea is using multi-photon scattering to deflect an atomic beam and separate them spatially \cite{Bernhardt1976, Li1983}. But because the photon recoil energy is quite small, hundreds of photons are needed to be scattered for atoms to gain enough momentum for separation , which will ultimately limit the isotope separation production rate. In contrast, optical pumping is a really efficient way to address isotopes. Atoms can be optical pumped to the desired states with almost single photon. Then they can be separated by magnetic gradient under the Stern-Gerlach scheme \cite{XW1992}. In order to effectively separate atoms with different magnetic moment in space, a high magnetic gradient field is required.  Nowadays, the widely used strong magnets, such as NdFeB magnet,  can provide strong magnetic field with a low cost.  The magnetic gardient can be created with magnets arranged in a hexaploe or quadrople shape \cite{Jerkins2010}. Recently, a Halbach-type magnet is introduced to deflect isotopes, which provides a convenient and low cost way to creating a high magnetic gradient \cite{Raizen2012, Mazur2014}. With those, a high isotope enrichment has been realized for Li atoms  \cite{Raizen2012}.

Here we study this magneto-optical combined method for K isotope separation. Potassium element has three isotopes with the natural abundance of 93.26\% for $^{39}$K, 6.73\% for $^{41}$K and  0.012\% for $^{40}$K. In cold atom physics, $^{40}$K is the first Fermion has been cooled down to the quantum degeneracy \cite{DeMarco1999}, and is still being extensively studied \cite{Haller2015,Cheuk2015,Nichols383}. Enriched $^{40}$K is an indispensable part of these experiments. Right now, the enriched $^{40}$K is usually obtained from the nuclear reaction, and the supply is not very stable. Additional convenient way to produce enriched $^{40}$K will be very useful for cold atom physics. 

Figure \ref{fig1} shows the energy level of K for different isotopes \cite{Kdata}. The isotope shifts are about few hundreds of MHz, which are much larger than the typical laser linewidth (about 1 MHz) and the upper state linewidth (about 6 MHz). Frequency locked diode lasers are used to perform optical pumping and fluorescence imaging. The lasers are locked to the K atomic transitions, and the laser linewidth is about 1 MHz.   $L_1$ and $L_2$ are pumping lasers for $^{39}$K and $^{41}$K respectively. Their frequencies are shown in Fig. \ref{fig1}, and are used to pump atoms from F=2 (low-filed seeking) state  to F=1 (high-filed seeking) state. Two lasers are combined and amplified by a tapered-amplifier, then coupled into a fiber and finally sent into the vacuum chamber with 10 mW for $L_1$ and 2 mW for $L_2$.  Both beams have the same beam size of 16 mm.    $L_P$ is the probe laser, and the frequency can be scanned over 1 GHz, covering the D1 transitions of the three isotopes (including $L1$ for $^{39}$K, $L2$ for $^{41}$K ). The beam size is 3 mm and intensity is about 2.8 mW/cm$^2$. The fluorescence induced by the probe beam are counted by a camera, and from this signal we can extract the flow rate of the atomic beam for different isotopes.

Figure \ref{fig2} (a) shows the experimental setup. The whole system is in a vacuum environment. It reachs a vacuum pressure of less than $10^{-5}$ Pa when the Potassium oven is not heated. The oven has a volume of {14 mL},  and is designed to be heated up to 600$^{\circ}$C. Limited by the deflecting magnets and also the transverse Doppler shift, the diverge angle of atomic beam should be small.  But small diverge angle means small flow rate. In order to get a high flow rate with a relatively small diverge angle, we use the micro-capillary array to collimate the atomic beams \cite{Senaratne2015}. As shown in Fig. \ref{fig2} (b). The needle has an inner diameter of 0.16 mm, and 8 mm long. After another collimation slit before the optical pumping, a  diverge angle of 2 degree is achieved. 

Out of the oven,  atoms are optical pumped to desired states. In order to reduce the Doppler shift,  both the optical pumping beams and probe beam are sent in  perpendicular to the moving direction of the atomic beam.   Then K atoms enter the magnetic region, where the magnets are aligned in Halbach style \cite{Mazur2014}. We used about 500 NdFeB permanent magnets (N52) to form a 1.5 meter long Halbach array, each one is $3\times 3\times 40$ mm. The maximum gradient of the magnetic field is about {2.5 $T/$cm} near the magnet surface (please see the supplementary material for more details). In this regime, high-field seeking atoms will move towards the magnets, hit the surface and get lost. Only low field seeking atoms will be deflected.

\begin{figure}[tbp]
\includegraphics[width=0.48\textwidth]{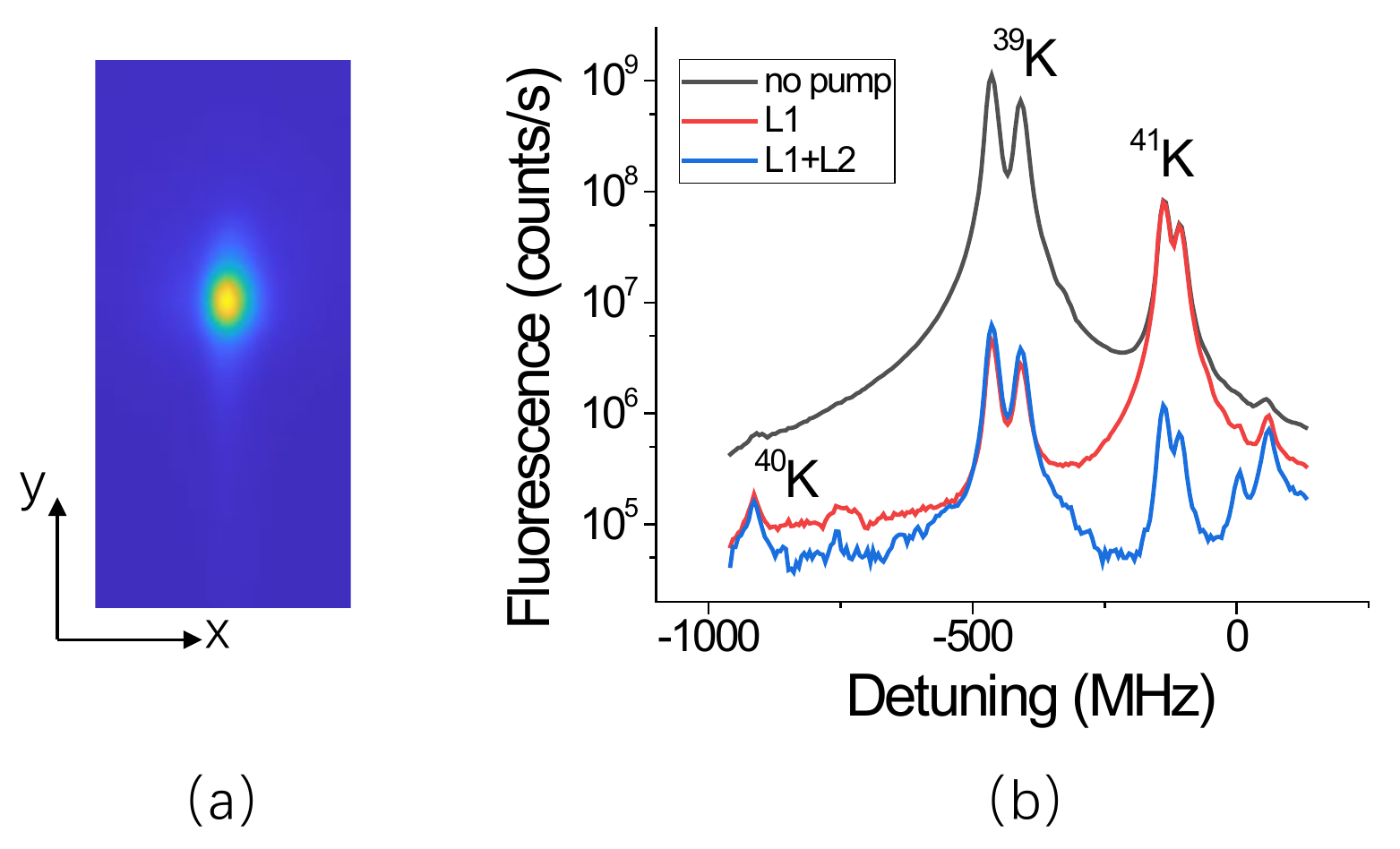}
\caption{\label{fig3}(a) A typical image of the atomic fluorescence recorded by the camera. (b) Fluorescent spectrum of potassium at 280$^{\circ}$C. The black line is the lineshape without optical pumping, where $^{39}$K peak value and $^{41}$K peak value have a ratio of $14:1$. The red line is with pumping light $L_1$ on. The $^{39}$K fluorescence is suppressed by more than two orders of magnitude , and the peak of $^{40}$K emerges. The blue line is with both pumping lights $L_1$ and $L_2$ on.  }
\end{figure}

Figure  \ref{fig2} (c) shows a typical simulation result of the atomic trajectory, which illustrate how the atoms are deflected by the magnet array. And finally, the deflected atoms are probed by a probe beam and the fluorescence is recorded by a camera (CCD).  Figure \ref{fig3} (a) shows a typical image of the deflected atoms. From the fluorescent image we can extract the total photon counts which is proportional to the atomic number and give a quantitative analysis of the isotope throughput.

By scanning the frequency of the probe laser, a high resolution Doppler-free spectrum is obtained. Figure \ref{fig3} (b) shows the typical data when the K oven is heated to 280$^{\circ}$C.  The frequency is scanned over the D1 line transitions of three isotopes, and the relevant transitions are identified. The black line shows the probe lineshape without pumping. The hyperfine structure is clearly show up and the peak value of $^{39}$K and $^{41}$K have a ratio of $14:1$, which is consistent with the natural abundance. But $^{40}$K can not be seen because of the high background of $^{39}$K fluorescence. The red line shows the lineshape when only $L_1$ pumping is on. Fluorescence of $^{39}$K is suppressed by more than two orders of magnitude. But for $^{41}$K, it almost has no change. At this time, because $^{39}$K is highly suppressed, the peaks of $^{40}$K shows up. The blue line shows the lineshape when both $L_1$ and $L_2$ are on. Then $^{41}$K is also suppressed about two orders of magnitude. From the data, we can see the abundance of $^{40}$K is about $2\%$, enhanced by two orders of magnitude.  To get a quantitative analysis of the experimental result, we define a suppression factor $\gamma=N_0/ N_1$, where $N_0$ and $N_1$ are the atom flow rate (fluorescence counts) before and after enrichment.

\begin{figure}[tbp]
\includegraphics[width=0.49\textwidth]{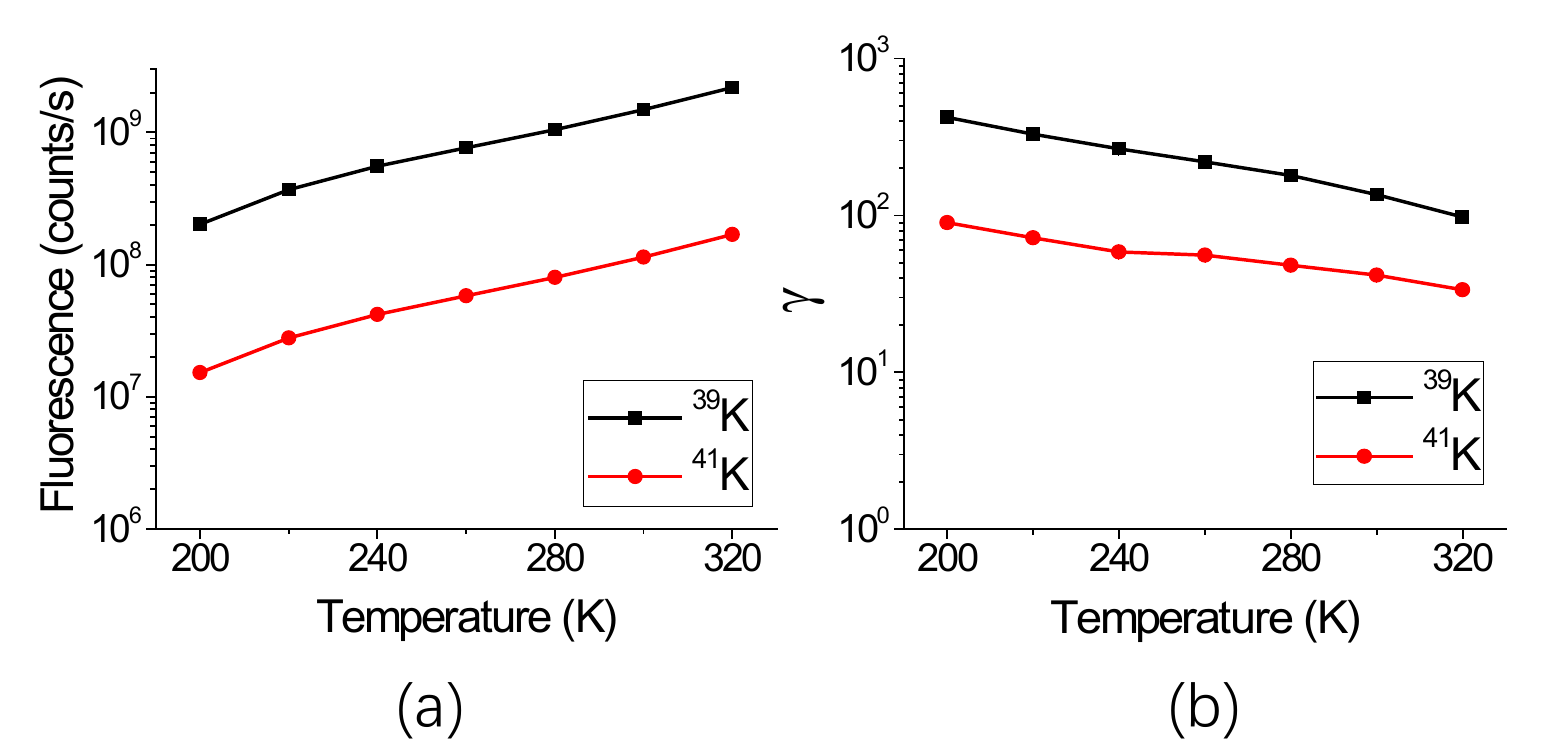}
\caption{\label{fig4}(a)The temperature dependences of atomic fluorescence. The fluorescence intensity increases 11 times from 200$^{\circ}$C to 320$^{\circ}$C. (b) The temperature dependences of  suppression factor. The supression factor decreases as the temperature increases.}
\end{figure}

Another important issue of isotope separation is the production rate. One effective way to increase the production rate is to increase the oven temperature. Figure \ref{fig4} (a) shows the fluorescence (thus the relative flow rate) of the atomic beam when changing the temperature. In the measured temperature regime, the flow rate increases $50\%$  for each 20 degrees. Figure \ref{fig4} (b) shows the suppression factor at different temperature. The total suppression factor is about two orders of magnitude, but decreases when the temperature increased. We contribute this to the reabsorption of the fluorescence. At the current temperature regime, the atomic density is {$1.9\times 10^{15}$ atoms/m$^3$}, and the collision between atoms is negligible \cite{Rosenberg1939}.  The fluorescence of the pumping process gives residual light in the chamber and will be reabsorbed by atoms. This will reduce the suppression rate when atom density get higher. 

With this data, we can make an estimation of the production rate. At 320$^{\circ}$C, with the pumping lights off, we capture the atomic fluorescence of $2.2\times10^9$ counts/s at $^{39}$K D1-line transition. Assuming one atom scatters one photon on average in the probe region, with a 0.005$\times4\pi$ solid angle, $80\%$ transmission rate (the loss mainly comes from uncoated lenses) and $50\%$ quantum efficiency of the camera, about $1.1\times10^{12}$ atoms pass the probe light per-second. Because the height of the magnet array is 14 times of the probe beam width, $1.54\times10^{13}$ atoms passes the magnet array in total, which means the apparatus defletes 0.086 mg $^{39}$K per-day. Accordingly, with the pumping lights on, the production rate of refined $^{40}$K is about 1.1 $\upmu$g/day. 

{\it Conclusion:---}
Combined the effective optical pumping and magnetic deflection, we show an enrichment of $^{40}$K by two orders of magnitude. This method can also be easily extended to get pure $^{39}$K or $^{41}$K. All we need to do is to shift the frequency of the pumping laser to other isotopes. Right now, the enrichment factor is limited by the imperfection of optical pumping, such as the reabsorption of atoms and residual light from the windows. These stray lights transmit around the vacuum chamber and pump the atoms to the opposite ground states out of the pumping region.

In the furture, we plan to improve our system. With the help of careful shielding and better pumping effect, 3 orders of magnitude suppression rate could be reached. The  micro-capillary oven can be optimized and then the oven temperature can be increased. With these, we should be able to increase the production rate by two order of magnitude, and reach 1 $m$g/week, which could be able to satisfy some realistic requirements, such as cold atom experiment with $^{40}$K.

This magneto-optical method for isotope separation can be easily extended to other species. For those isotopes whose ground states already have nonzero spin or orbital momentum, half of the ground states will be high-field seeking states, and half of the states will be low-field seeking states. This isotope separation method can be directly applied if suitable lasers are available. For those isotopes whose ground states have zero magnetic moment (including Alkaline-earth metals, Zinc, Mercury and Ytterbium), multi-laser system also exists, which is able to drive atoms from the ground states to the metastable states with particular magnetic moment \cite{Mazur2014}.

{\it Acknowledgement:---}
We acknowledge the support from the National Key R\&D Program of China under Grant No.2018YFA0307200, the National Natural Science Foundation of China under Grant No. 12074337, Natural Science Foundation of Zhejiang province under Grant No. LR21A040002, LZ18A040001, Zhejiang Province Plan for Science and technology No. 2020C01019 and the Fundamental Research Funds for the Central Universities under No. 2020XZZX002-05 and 2021FZZX001-02 .

\bibliography{Kref1.bib}

\end{document}


\newcommand{\smtitle}[1]{\begin{center}\large{\textbf{#1}}\end{center}\vskip 2em}
\renewcommand{\thesection}{\Alph{section}}
\renewcommand{\thefigure}{S\arabic{figure}}
\renewcommand{\thetable}{S\Roman{table}}
\setcounter{figure}{0}
\renewcommand{\theequation}{S\arabic{equation}}
\setcounter{equation}{0}

\appendix
\begin{widetext}
\smtitle{Supplementary Material for ``Isotope Separation of Potassium with a magneto-optical combined method''}

Here we provides some details about the shape of the magnet array, and the collision rate between atoms in the atomic beam.

\section{ Magnetic Filed Geometry}

As shown in Fig. S1, the Halbach array we used in this experiment has a curved surface depicted by a function $L(\theta)=Le^{\theta/\tan\alpha}$. Different from a straight array, every point on the curve has the same intersection angle with the origin point (atomic source), guaranteeing a constant incident angle between the magnet surface and the isotopes.  To deflect a high speed low-filed seeking atom, we need to make sure that the velocity component of the atom perpendicular to the magnets will decrease to zero before touching the magnets surface.
$$v\sin{\alpha}<\sqrt{2\mu_B\Delta B/m}$$
While a smaller incident angle permits a larger speed range of the atoms,  the length of the magnets array grows rapidly. An appropriate constant angle helps us to maximize the reflection of the isotopes with a limited length of the vacuum chamber (2 meter approximately).
\\

\begin{figure}[hbp]
\includegraphics[width=0.9\textwidth]{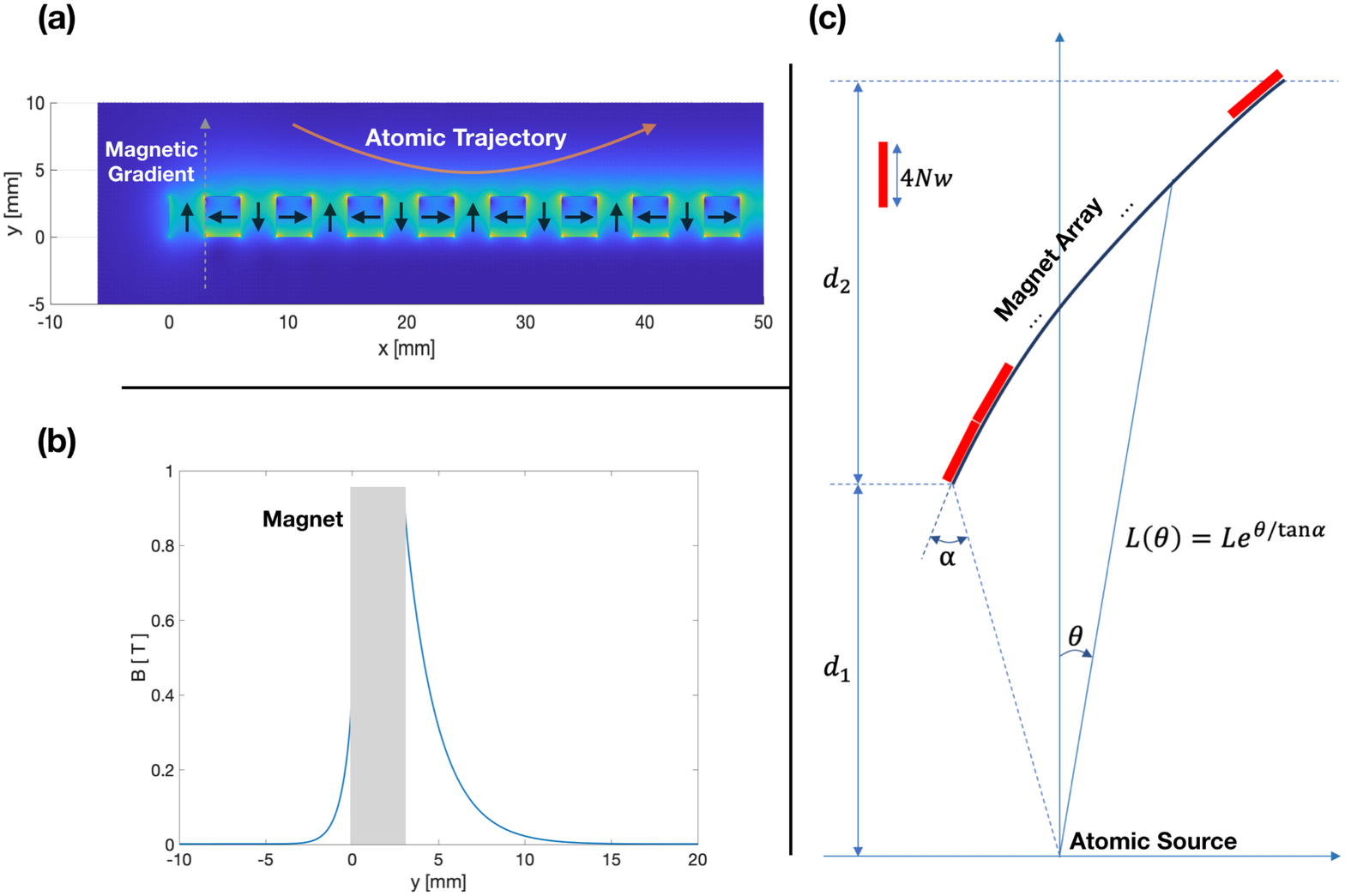}
\caption{\label{fig1}Characteristics of the magnetic field. (a) A typical Halbach array and the spatial distribution of the magnetic flux density. (b) Magnetic gradient of a Halbach array. (c) Curved magnets array formed by 15 short flat arrays. Each short array has 8 periods and 32 pieces of N52 magnets which is 3mm wide. The total length of the curved array is 1.5 m. The curve is depicted by $L(\theta)=Le^{\theta/\tan\alpha}$. Every point on the curve has the same intersection angle $\alpha$ with the origin point. $d_1=0.7$ m, $d_2=1.5$ m, $\alpha=0.02$ rad, $N=8$, $w=3$ mm. }
\end{figure}

\section{Particle Collision and Relaxation}

The mean free path of an ideal gas is a function of  particle density $n$ and particle diameter $d$,
$$\lambda=\frac{1}{\sqrt{2}\pi d^2n}$$ 
Although this formula is only strictly valid for an isotropic gas model, we can utilize it to get a lower bound of the mean free path of our system.
We estimate the atomic density through pumping region by measuring the absorption of a probe light (with power of 0.1mW and a beam waist of 3 mm) at $^{39}$K $F=2\rightarrow F'=1$ transition. A 11$\mu$W absorption corresponds to an atomic flux close to $4.3\times 10^{13}$ atoms per second. Scaling to the whole aperture (10 mm height), the total flux is $1.4\times 10^{14}$ atoms per second. Dividing the total flux by the product of average velocity and aperture area, the atomic density is $1.9\times 10^{15}$atoms/m$^3$ and the mean free path is 386 m. 
In the 2 meter long vacuum chamber used in our experiment, particles rarely have collision or relaxation processes. This keeps their quantum state unchanged and enables us to analysis isotope separation effect and the throughput of the apparatus by a single particle model. As we discussed in the main text before, We attribute the decrease of the suppression factor at high temperature to the imperfection of optical pumping and reabsorption of the fluorescence, which can be improved by extending the length of the pumping region and carefully shielding the residual light reflected by the chamber.
\end{widetext}